\documentclass[useAMS,usenatbib]{mn2e}
\usepackage{color,url,graphicx, amsmath}
\usepackage{epstopdf}

\title[A Clean Sightline to Quiescence]{A Clean  Sightline to Quiescence:  Multiwavelength Observations of  the High Galactic Latitude Black Hole X-ray Binary Swift J1357.2-0933}
\author[Plotkin et al.]{%
	Richard~M.~Plotkin,$^{1,2}$\thanks{E-mail: richard.plotkin@curtin.edu.au}
         Elena Gallo,$^{1}$
	Peter G. Jonker,$^{3,4}$
	James C.~A.~Miller-Jones,$^2$
	\newauthor
	Jeroen Homan,$^5$  
	Teo Mu\~{n}oz-Darias,$^{6,7}$
	Sera Markoff,$^8$ 
	Montserrat Armas Padilla,$^{6,7}$
	\newauthor
	Rob Fender,$^{9}$
	Anthony~P.~Rushton,$^{9,10}$
	David~M.~Russell,$^{11}$
	and Manuel A.~P.~Torres$^{3,4,12}$
	\\
$^{1}$Department of Astronomy, University of Michigan, 1085 South University Ave, Ann Arbor, MI 48109, USA\\
$^2$International Centre for Radio Astronomy Research, Curtin University, G.P.O. Box U1987, Perth, WA 6845, Australia\\
$^3$SRON, Netherlands Institute for Space Research, Sorbonnelaan 2, 3584-CA, Utrecht, The Netherlands\\
$^4$Department of Astrophysics/IMAPP, Radboud University Nijmegen, Heyendaalseweg 135, 6525-AJ, Nijmegen, The Netherlands\\
$^5$Kavli Institute for Astrophysics and Space Research, Massachusetts Institute of Technology, 70 Vassar Street, Cambridge, MA 02139, USA\\
$^{6}$Instituto de Astrof'sica de Canarias, E-38205 La Laguna, Tenerife, Spain\\
$^7$Departamento de astrof'sica, Univ. de La Laguna, E-38206 La Laguna, Tenerife, Spain \\
$^{8}$Anton Pannekoek Institute for Astronomy, University of Amsterdam, Science Park 904, 1098 XH, Amsterdam, the Netherlands\\
$^{9}$Department of Physics, Astrophysics, University of Oxford, Keble Road, Oxford OX1 3RH, UK\\
$^{10}$School of Physics and Astronomy, University of Southampton, Highfield, Southampton SO17 1BJ, UK\\
$^{11}$New York University Abu Dhabi, PO Box 129188, Abu Dhabi, UAE\\
$^{12}$European Southern Observatory, Alonso de C\'{o}rdova 3107, Vitacura, Casilla 19001, Santiago de Chile, Chile
}

\newcommand{\aj}{AJ}
\newcommand{\mnras}{MNRAS}
\newcommand{\apjs}{ApJS}
\newcommand{\apj}{ApJ}
\newcommand{\apjl}{ApJL}
\newcommand{\araa}{AR\&A}

\newcommand{\aap}{A\&A}

\newcommand{\ssr}{Space Science Reviews}

\def\lesssim{\mathrel{\hbox{\rlap{\hbox{\lower3pt\hbox{$\sim$}}}\hbox{\raise2pt\hbox{$<$}}}}}
\def\gtrsim{\mathrel{\hbox{\rlap{\hbox{\lower3pt\hbox{$\sim$}}}\hbox{\raise2pt\hbox{$>$}}}}}

\newcommand{\fullsrc}{Swift J1357.2$-$0933}
\newcommand{\src}{J1357.2}
\newcommand{\srcasix}{A0620$-$00}
\newcommand{\srcxte}{XTE~J1118$+$480}

\newcommand{\srcvfour}{V404~Cyg}
\newcommand{\lr}{L_r}
\newcommand{\lx}{L_X}

\newcommand{\lledd}{L_X/L_{\rm Edd}}
\newcommand{\ledd}{L_{\rm Edd}}
\newcommand{\mbh}{M_{\rm BH}}
\newcommand{\msun}{{\rm M_{\sun}}}

\newcommand{\ergs}{{\rm erg~s}^{-1}}

\newcommand{\flux}{{\rm erg~s}^{-1}~{\rm cm^{-2}}}
\newcommand{\cmtwo}{{\rm cm}^{-2}}
\newcommand{\rg}{r_{\rm g}}

\newcommand{\xrb}{BHXB}
\newcommand{\nh}{N_{\rm H}}

\begin{document}
\date{}
\pagerange{\pageref{firstpage}--\pageref{lastpage}} \pubyear{}
\maketitle
\label{firstpage}

\begin{abstract}
We present coordinated multiwavelength observations of the high Galactic latitude (b=+50$^\circ$) black hole X-ray binary (\xrb) \fullsrc\ in quiescence. Our broadband spectrum includes  strictly simultaneous radio and X-ray observations, and  near-infrared, optical, and ultraviolet data taken 1-2 days later.  We detect \fullsrc\  at all wavebands except for the radio ($f_{\rm 5GHz}$$<$$3.9\ \mu$Jy beam$^{-1}$; 3$\sigma_{\rm rms}$).  Given  current constraints on the  distance (2.3-6.3~kpc), its 0.5-10~keV X-ray flux corresponds to an Eddington ratio  $\lledd = 4\times10^{-9}-3\times10^{-8}$ (assuming a black hole mass of 10$\msun$).   The broadband spectrum is dominated by synchrotron radiation from a  relativistic population of outflowing thermal electrons, which we argue to be a common  signature of short-period quiescent \xrb s.  Furthermore, we identify the frequency where  the synchrotron radiation transitions from optically thick-to-thin ($\nu_b \approx2-5\times10^{14}$~Hz, which is the most robust determination of a `jet break' for a quiescent \xrb\ to date.   Our interpretation relies  on the  presence of steep curvature in the ultraviolet spectrum, a frequency window made observable by the  low amount of interstellar absorption along the line of sight.  High Galactic latitude systems like \fullsrc\  with clean ultraviolet sightlines are crucial  for understanding black hole accretion at low luminosities.  
\end{abstract}


\section{Introduction}
\label{sec:intro}
   Black hole X-ray binaries  (\xrb s) in the hard X-ray spectral state (see  \citealt{remillard06} for a review) are nearly always associated with compact radio emission  from a steady state relativistic jet \citep[e.g.,][]{hjellming88, fender01}.     Correlated radio and X-ray variability  on day- to week-long timescales implies a coupling between the relativistic  outflow and the underlying accretion flow \citep[e.g.,][]{heinz03, markoff03, corbel13, gallo12, gallo14}.  The radio emission arises from partially  self-absorbed synchrotron radiation from the compact jet \citep{blandford79}.  Meanwhile,  X-ray emission includes contributions from a radiatively inefficient accretion flow \citep[RIAF; e.g.,][]{ichimaru77, narayan94, abramowicz95,blandford99,narayan00,quataert00, yuan05} and optically thin   synchrotron radiation from the jet \citep[e.g.,][]{markoff01, markoff03, russell10, russell13, plotkin12a}.   A cool inner disk  may also contribute X-ray emission \citep{miller06, reis10}.

A substantial  number of hard state \xrb s have been discovered to be radio-underluminous  at a given  X-ray luminosity, when compared to more `traditional' systems \citep[e.g.,][]{jonker04, corbel04, brocksopp05, cadolle-bel07, rodriguez07, xue07, coriat11, soleri11, gallo12, cao14, meyer-hofmeister14}.  It is not clear, however,  whether this  diversity in  radio properties continues after \xrb s transition into the quiescent spectral state \citep[Eddington ratios\footnote{The Eddington luminosity for hydrogen in a spherical geometry is $\ledd=1.26 \times 10^{38}\left(M/\msun\right)~\ergs$, where $M$ is the black hole mass.} $\lledd \lesssim 10^{-5}$;][]{plotkin13}.  In fact, at least three `radio-underluminous'  systems have been observed to transition to the `traditional' radio/X-ray luminosity correlation\footnote{We (somewhat arbitrarily) call \xrb s that follow radio/X-ray luminosity correlations of the form $\lr \propto \lx^{\approx 0.6}$ as `traditional' here for historical reasons \citep[e.g.,][]{gallo03}.} 
 around $\lledd \lesssim 10^{-4-5}$ (e.g., H1743$-$322, \citealt{jonker10, coriat11}; XTE~J1752$-$223, \citealt{ratti12}; MAXI~J1659$-$152, \citealt{jonker12}), perhaps hinting that the `radio-underluminous' \xrb\ branch does not extend indefinitely toward the lowest X-ray luminosities.  
 
Our currently limited knowledge  on quiescent \xrb\ accretion flows/jets stems primarily from  their  low luminosities levels, combined with the small number of  known \xrb s located close  to the Earth  \citep[e.g.,][]{calvelo10, miller-jones11}.   There are currently only three  low-mass  \xrb\ systems (with a confirmed black hole accretor) that have meaningful, \textit{simultaneous} radio and X-ray    constraints on their jets in quiescence, \srcvfour\ ($\lledd \approx 10^{-6}$; \citealt{hjellming00, gallo05, hynes09}), \srcasix\ ($\lledd \approx 10^{-8.5}$; \citealt{gallo06}), and \srcxte\ ($\lledd \approx 10^{-8.5}$; \citealt{gallo14}).  The (high-mass) Be/black hole X-ray binary system MWC 656 also has non-simultaneous radio and X-ray detections in quiescence ($\lledd \approx 10^{-8}$; \citealt{munar-adrover14, dzib15}).  Multiwavelength constraints on more  systems are needed to understand the disk/jet connection in quiescence.

Here, we present new coordinated radio, near-infrared (NIR), optical, ultraviolet (UV) and X-ray observations of the \xrb\ \fullsrc\ (hereafter \src) in quiescence.    In Section \ref{sec:src}, we provide a brief overview of the properties of \src.  This source displayed unusual behavior in the hard state \citep{corral-santana13}, and Section \ref{sec:src} is intended to help put the current work into  context.  In Section~\ref{sec:obs} we describe our multiwavlength observations, and  results are presented in Section~\ref{sec:res}.  In Section~\ref{sec:wisedisc} we describe a serendipitous discovery of the 2011 outburst from  archival infrared data.  Finally, our results are discussed in Section~\ref{sec:disc}, and our main conclusions are highlighted in Section~\ref{sec:conc}.  Error bars are quoted at the 68\% confidence level, unless stated otherwise. We adopt a distance ranging from 2.3-6.3~kpc (\citealt{shahbaz13, mata-sanchez15}),  an orbital inclination angle $i>70^\circ$ \citep{corral-santana13, torres15}, and we assume a black hole mass of 10~$\msun$ \citep{mata-sanchez15}.

\section{\fullsrc}
\label{sec:src}
\src\ was discovered by the Swift Burst Array Telescope (BAT) on 2011 Jan 28  \citep{krimm11}.   The distance is not well determined, and we adopt a  range of $2.3<d<6.3$ kpc here.  The lower limit is based on disk veiling constraints from optical spectroscopy in quiescence \citep{mata-sanchez15}; the upper limit arises from an estimate of the degree to which  synchrotron radiation  could be diluting starlight from the companion, given the orbital period and inclination of the system \citep{shahbaz13}.    If the  distance happens to fall toward the lower end, then \src\ would be the least-luminous known \xrb\ in quiescence \citep{armas-padilla14a}, and it would be one of the few known \xrb s suitable for  deep radio observations to search for a quiescent radio jet.   Furthermore, \src\   has a high Galactic latitude ($b=+50^\circ$), making it one of the few known \xrb s with a low-enough Galactic absorption column density to allow  UV studies  in quiescence (\srcxte\ is another notable \xrb\ at high Galactic latitude; see \citealt{mcclintock03}).   

 From time-resolved optical spectroscopy of broad, double-peaked H$\alpha$ emission in outburst, \citet{corral-santana13} found  an orbital period of $2.8\pm0.3$ h, and \citet{mata-sanchez15} recently constrained the black hole mass to be  $>$9.3~$\msun$.  \citet{corral-santana13} also discovered recurring dips in the  outburst optical light curve on 2-8 minute timescales, where the flux dropped by up to $\sim$0.8~mag.      They explain this remarkable short-term variability as \src\ being viewed at a nearly edge-on inclination angle ($i\gtrsim 70^\circ$), with the dipping behavior  being caused by a geometrically thick obscuring torus in the inner disk.   From time-resolved spectroscopy of the quiescent optical counterpart,  \citet{torres15} also favor a high orbital inclination, based on the profiles of broad, double peaked H$\alpha$.   The proposed geometrically thick inner toroidal structure could be a crucial component of all accretion flows (and potentially important for producing and sustaining jets), but its signatures cannot be detected in other systems at lower inclination angles.  

The odd optical `dipping' behavior observed during outburst  persists into quiescence, albeit at longer recurring timescales of $\sim$30-m \citep{shahbaz13}.  Intriguingly, \citet{shahbaz13}  find that, superposed on the dips, there is stochastic, large-amplitude optical/NIR variability  (the fractional optical rms  is  $\approx$35\%).  This variability is highlighted by 10-30~m flare events with amplitudes up to 1.5-2~mag.  Due to this variability, as well as a steep NIR/optical spectrum ($f_\nu \propto \nu^{-1.4}$), \citet{shahbaz13} argue that the NIR/optical spectrum is dominated by  synchrotron radiation emitted by a thermal jet.

\section{Observations and Data Reduction}
\label{sec:obs}
We targeted \src\ through a joint \textit{Chandra}-NRAO program (ObsID 15782;  PI Plotkin), with simultaneous Chandra and VLA observations   taken on 2014 March 20 (see Section \ref{sec:obs:simul}).  We also obtained quasi-simultaneous NIR, optical, and UV observations, which  were  taken 1-2 days after the radio/X-ray data (see Section~\ref{sec:obs:quasi}).  The full spectral energy distribution (SED) is summarized in Table~\ref{tab:sed}.

\begin{table*}
\begin{minipage}{0.9\linewidth}
\begin{center}

\caption{Observing Log and SED}
\label{tab:sed}

\begin{tabular}{l l l l l l }
\hline
		Telescope & 
		Date\footnote{The universal time (UT) is listed for the beginning of the simultaneous radio and X-ray observations.   For the NIR and optical observations, data were taken by cycling through each filter over each observing night.  The Swift  observations were taken by cycling through each UVOT filter (with 60-360 s exposures), beginning at UT 07:40.}       & 
		Filter & 
		Exposure Time & 
	 	Frequency &  
		Flux Dens.\footnote{Flux densities are reported after applying corrections for interstellar extinction, as described in Section \ref{sec:obs:quasi}.  See the text for values  prior to applying  extinction corrections.  Upper limits are quoted at the 3$\sigma$ level, and all error bars are  at the 68\% level.  Upper limits for VLA radio observations have units $\mu$Jy beam$^{-1}$.} \\ 

	                        & 
	                        & 
	                        & 
	            (ks)            & 
	           (Hz)     & 
	           ($\mu$Jy)  \\ 
		\hline
		
		 VLA & 2014 March 20  UT 04:20  &    C-band  &   30.5  & $5.30\times10^{9}$   & $<$4.2 \\
		 VLA\footnote{Throughout the text, we adopt the more sensitive radio upper limit obtained when combining the 2014 radio observation with two archival VLA observations from 2013 (see Section~\ref{sec:obs:radio}).}   & 2013 July 9-11 + 2014 March 20  & C-band &  46.2 & $5.30\times10^9$  & $<$3.9 \\
		 WHT  & 2014 March 20-21   &      $K_S$       &   5     & $1.39\times10^{14}$  & $32.2 \pm 1.6$ \\
		 WHT   & ...    			       &   $H$           &  5        & $1.80\times10^{14}$  & $30.6 \pm 1.3$ \\
		 WHT    & ...		                &   $J$            &   4.5     & $2.43\times10^{14}$  & $29.8 \pm 1.0$ \\
		 LT        & 2014 March 21-22    &  $i^\prime$    &  1.4       & $4.01\times10^{14}$ & $28.4 \pm 1.0$ \\
		 LT         & ...			       &     $r^\prime$   &   1.6       & $4.86\times10^{14}$ & $22.7 \pm 1.0$ \\
	  	  \textit{Swift}    &  2014 March 21          & $u$  & 1.7   &    $8.65\times10^{14}$ & $12.4 \pm 1.8$ \\
		  \textit{Swift}    &  ...		                & $uvw1$ & 1.7 &   $1.15\times10^{15}$ & $3.1 \pm 1.0$ \\
		\textit{Swift}    &  ...		                &   $uvm2$ & 1.7  &   $1.33\times10^{15}$ & $<$2.1 \\
		 \textit{Chandra} & 2014 March 20  UT   02:40    & ACIS-S3 & 25.5  & 0.5-10~keV  &  $\left(8.2\pm1.9\right)\times10^{-15}$\ \footnote{ Unabsorbed flux  from 0.5-10~keV in erg~s$^{-1}$~cm$^{-2}$, assuming $\Gamma=2.1$ and no intrinsic absorption.} \\

\hline
\end{tabular}
\end{center}
\end{minipage}
\end{table*}

\subsection{Strictly Simultaneous Observations}
\label{sec:obs:simul}

\subsubsection{Radio Observations}
\label{sec:obs:radio}
The radio observations were taken with the Karl G. Jansky Very Large Array (VLA) on 2014 March 20, from 04:20--14:18 UT under project code SF0459.  The array was in its most-extended A-configuration.   We observed in two 1024-MHz basebands, with central frequencies of 4.8 and 5.8\,GHz.  The data were processed using the Common Astronomy Software Application \citep[CASA;][]{mcmullin07}.  We used 3C\,286 to set the amplitude scale according to the \citet{perley13} coefficients within CASA's {\sc setjy} task, and we used the extragalactic calibrator source J1408-0752 to determine the complex gain solutions in the direction of the target.  Our on-source time was 508\,min.  Following external gain calibration, we made an image of the field surrounding \src, using two Taylor terms to model the frequency dependence of the sources in the field and thereby avoid amplitude errors in the deconvolution.  We used Briggs weighting with a robust value of 1 as the best compromise between achieving high sensitivity and downweighting the sidelobes of the dirty beam.  We placed outlier fields on known bright sources outside the main image, to ensure that their sidelobes did not affect the final image.  We do not detect \src\ to a $3\sigma$ upper limit of 4.2\,$\mu$Jy\,beam$^{-1}$.  

In an attempt to place a deeper radio flux density limit, we also retrieved two archival VLA observations taken under project code 13A-203 (PI Fender), which  added   261 min on-source.  The two archival observations were taken on 2013 July 9--10 (23:13-01:57 UT) and 10--11 (23:24--02:08 UT), when the array was in the more compact C-configuration.  The observing setup and calibration sources were  identical to our 2014 observation.  We calibrated the two archival data sets separately, and we combined all three epochs of VLA data into a single image.   When creating the combined image, we  accounted for the mismatch in angular scales probed by the two array configurations by using the multi-scale clean algorithm implemented in CASA's {\sc clean} task, and we tried different data weighting schemes.  Our best image was made by removing the shortest baselines ($<$10\,k$\lambda$) and using a robust weighting of 1.  \src\ is not detected in the combined image, with a  3$\sigma$ upper limit of 3.9\,$\mu$Jy\,beam$^{-1}$.   We adopt this more sensitive $<$3.9\, $\mu$Jy\,beam$^{-1}$ limit throughout the text, which corresponds to a radio luminosity $\lr<1.3\times10^{26} - 9.8\times10^{26}~\ergs$ at 5.3~GHz, assuming a flat radio spectrum and $2.3<d<6.3$~kpc.

\subsubsection{X-ray}
\label{sec:obs:xray}
The Chandra observations were  taken on 2014 March 20 UT 02:40--10:40.    \src\ was placed at the aimpoint of the S3 chip on the Advanced CCD Imaging Spectrometer  \citep[ACIS;][]{garmire03}.  Data were telemetered in very faint mode, which we then  reprocessed with the \textit{Chandra} Interactive Analysis of Observations  software  \citep[CIAO;][]{fruscione06}, applying the latest calibration files (CALDB 4.6.5).  We removed 200s from the exposure, during which there was a slightly elevated sky level, yielding an effective exposure time of 25.5 ks.  The remaining  analysis was performed over 0.5-7~keV.  Photometry was performed over a circular aperture with a 10 pixel radius, centered at the optical/NIR position from \citet{rau11}.  The background was estimated over a circular annulus with inner and outer radii of 20 and 40 pixels, respectively.   We obtained 21 total counts, with an estimated 1.5 background counts in the circular aperture, yielding a net count rate of $\left(0.8 \pm 0.2\right) \times 10^{-3}$ counts s$^{-1}$.   We use the Interactive Spectral Interpretation System \citep[ISIS;][]{houck00} to calculate an unabsorbed 0.5-10~keV model flux of   $f_X= \left(8.2\pm1.9\right)\times10^{-15}~\flux$, assuming an unabsorbed power-law\footnote{$N_E = N_0 \left(E/E_0\right)^{-\Gamma}$, where $\Gamma$ is the X-ray photon index,  $N_E$ is the photon flux density at energy $E$, and $N_0$ is the normalization at energy $E_0=1$~keV.} 
with $\Gamma=2.1$ (see Section~\ref{sec:xrayspec}).   For distances 2.3-6.3~kpc, our 2014  \textit{Chandra} flux corresponds to 0.5-10~keV luminosities $5.2\times10^{30} - 3.9\times10^{31}~\ergs$ and  $\lledd = 4.1\times10^{-9}-3.1\times10^{-8}$ (assuming $\mbh=10~\msun$).

\subsection{Quasi-simultaneous Observations}
\label{sec:obs:quasi}

Observations  at other wavebands were taken 1-2 days after the simultaneous radio VLA and \textit{Chandra} X-ray observations.  All magnitudes and flux densities  in the following text  are reported prior to correcting for Galactic extinction; all  data  in  figures and in Table~\ref{tab:sed} are  presented after correcting for  extinction, assuming $A_V=0.123$ and $E(B-V)=0.04$ in the optical/UV \citep[as adopted by][]{armas-padilla13, shahbaz13}.  For  UV observations with the Swift   Ultraviolet/Optical Telescope \citep[UVOT;][]{roming05} (see Section~\ref{sec:obs:uv}), we adopt the $A_\lambda/A_V$ ratios  tabulated by \citet{kataoka08} for each UVOT filter. 

\subsubsection{Near-infrared}
\label{sec:obs:nir}
NIR observations were taken on the night starting 2014 Mar 20 (PI Jonker), using the Long-slit Intermediate Resolution Infrared Spectrometer (LIRIS) on the 4.2 m William Herschel Telescope (WHT).  We cycled through the $J, H$, and $K_s$ filters over the night.  Seeing conditions were generally poor, approximately $1\farcs25$ (full width half max) in the $K_S$ band.   For the first sequence in the $J$ filter, we applied a 5 point dither pattern, taking 5 exposures of 20 s each per dither position.  Once we visually confirmed that \src\ was detected in that sequence, we switched to a 10-point dither pattern  for all filters (with 5$\times$20 s exposures per position).   We cycled through  5 dither sequences per filter over the night, exposing for  a total of 4500 s in the $J$ filter, and 5000 s in $H$ and $K_s$. 

 Sky subtraction and flat-fielding were perfomed using routines from the LIRIS data reduction pipeline {\tt THELI} \citep{schirmer13}.  Individual frames were then combined to create a single image in each filter.   \src\ was detected  in all three filters.  We performed differential photometry, calibrated to two nearby (unsaturated) stars detected in the Two Micron All Sky Survey \citep[2MASS;][]{skrutskie06}.  We find $J=19.32\pm0.03$, $H=18.81\pm0.04$ and $K_s=18.29\pm0.05$~mag (on the 2MASS magnitude scale), with the uncertainties dominated by the error in the 2MASS photometric zeropoints.   

We  searched for intranight NIR variability by  co-adding the exposures from each individual dither sequence, which resulted in five time-resolved images per filter.  Each time-resolved image represents 1000~s of exposure time, except for the first $J$-band image which includes 500~s.    We do not detect any obvious short-term variability that is significantly larger than the uncertainty on each flux measurement: the magnitudes in each sliced $J$, $H$, and $K_s$ image  vary by $\pm 0.16$, 0.14, and 0.06~mag, respectively, but uncertainties on each magnitude measurement are typically comparable, with  $\sigma_m \approx \pm 0.13$,  0.09, and 0.10~mag, respectively in each filter.   We are likely not sensitive to the short-term NIR variability observed by \citet{shahbaz13}, given the cadence and length of our NIR exposures. 

\subsubsection{Optical}
\label{sec:obs:opt}
We observed \src\ in the optical on the night starting 2014 Mar 21 (PI Fender) with the IO:O camera on the 2 m Liverpool Telescope on La Palma, Spain \citep{steele04}.  We alternated between the $r^\prime$ and $i^\prime$ filters, taking a total of 18$\times$200 s exposures in each filter. Observations were carried out under very variable weather and seeing conditions, especially during the first half of the run.  For this reason we only consider the last eight exposures taken in each band for our analysis (one additional $i^\prime$ exposure was excluded because of a cosmic ray).

Data were bias subtracted and flat-field corrected using standard procedures in {\tt IRAF}.  Photometry was calibrated toÊ nearby stars in the SDSS catalog.  We find $r^\prime$ and $i^\prime$ in the range $20.4\pm0.1$ -- $20.9\pm0.2$ and $20.1\pm0.1$ -- $20.7\pm0.2$~mags, respectively, consistent with the optical variability on minute-long timescales reported by \citet{corral-santana13} and \citet{shahbaz13}. In order to increase signal-to-noise (and to average over the short-term variability),Ê the individual exposures were combined to produce a single image per filter. This yields $r^\prime=20.62\pm0.05$ and $i^\prime=20.35\pm0.04$~mag after 1600 and 1400 s on target, respectively.Ê

\subsubsection{Ultraviolet}
\label{sec:obs:uv}
\vspace{-0.2cm}
UV observations were taken with Swift/UVOT on 2014 March 21 07:40 (PI Homan).  Data were taken by cycling through the  $u$, $uvw1$, and $uvm2$ filters (1700~s exposures in each filter).   Individual frames were combined using the {\tt uvotimsum} tool. Flux densities were obtained with the {\tt uvotsource} tool, using circular source and background extraction regions with 4\arcsec\ and 11\farcs5 radii, respectively.  \src\ was detected in the $u$ and $uvw1$ filters, with flux densities of $10.3\pm1.8$ ($u$) and $2.40\pm0.98$ $\mu$Jy ($uvw1$).  \src\ was not detected in the $uvm2$ filter, with a flux density $<$1.63 $\mu$Jy (3$\sigma$ limit). 

 \begin{figure}
\includegraphics[scale=0.48]{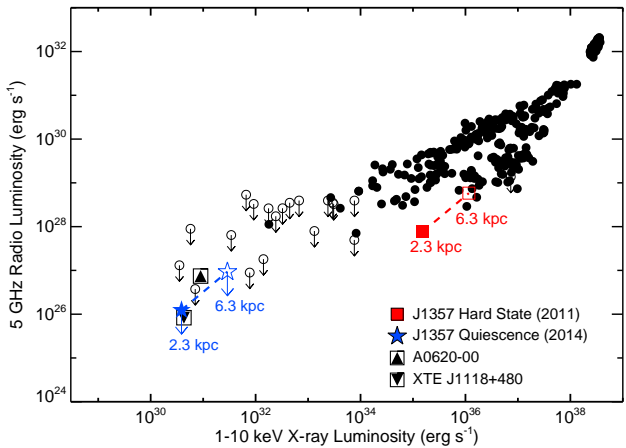}
 \caption{\src\ in the radio/X-ray luminosity plane, including our observations in quiescence (blue star symbols) and nearly simultaneous hard-state observations from the literature  (red squares).  The luminosity range due to the unknown distance of \src\ is illustrated by dashed lines connected to filled (2.3~kpc) and open (6.3~kpc) symbols.  Data points  for other \xrb s are taken from \citet{gallo14}, where filled circles denote radio detections, and open circles/arrows denote radio non-detections.  The locations of \srcasix\ and \srcxte\ in quiescence are highlighted in the figure (see legend; both \xrb s have radio detections at $\lledd\approx 10^{-8.5}$).  }
  \label{fig:lrlx}
 \end{figure}

\section{Results}
\label{sec:res}

\subsection{X-ray Spectrum}
\label{sec:xrayspec}
We extracted an X-ray spectrum using the CIAO tool {\tt specextract}.  Given the low-number of X-ray counts, we only attempted to fit a  {\tt powerlaw} model to the spectrum.  We perform the spectral fitting within ISIS, using  Cash statistics \citep{cash79} with the background included in the  fit (an energy-dependent aperture correction was applied to account for the finite size of the  extraction region).   We initially fixed the column density to the Galactic value of $\nh=1.2\times10^{20}~\cmtwo$ \citep{krimm11a, armas-padilla14a},  and we found a best-fit photon index  $\Gamma=2.6\pm0.9$.   \citet{armas-padilla14} found negligible X-ray absorption for \src\ from a high-count \textit{XMM-Newton} X-ray spectrum taken during its 2011 outburst.  Indeed, when we refit the quiescent \textit{Chandra} X-ray spectrum allowing $\nh$ to vary as a free parameter, $\nh$ converges toward zero and the best-fit photon index remains similar  ($\Gamma=2.6_{-0.8}^{+1.2}$). This photon index is consistent with an  \textit{XMM-Newton} observation in quiescence taken by \citet{armas-padilla14a} in 2013 July, who found a best-fit  $\Gamma=2.1\pm0.4$, and it is typical of other quiescent \xrb\ systems ($\Gamma \sim 2.1$; \citealt{plotkin13, reynolds14}).  To ease comparison to \citet{armas-padilla14a}, we adopt the  canonical $\Gamma=2.1$ and no column density for flux and luminosity calculations, unless stated otherwise.\footnote{Adopting $\Gamma=2.6$ and $\nh=1.2\times10^{20}$~cm$^{-2}$  provides an unabsorbed 0.5-10 keV model flux $f_x=\left(7.1\pm2.1\right)\times10^{-15}\ \flux$, which is similar within the errors   to the $f_X=\left(8.2\pm1.9\right)\times10^{-15}~\flux$ reported in Section~\ref{sec:obs:xray} when adopting $\Gamma=2.1$ and no absorption.  The difference in luminosity between the two spectral fits is therefore negligible  compared to the uncertainty on the  distance.}
   
 \begin{figure}
\includegraphics[scale=0.48]{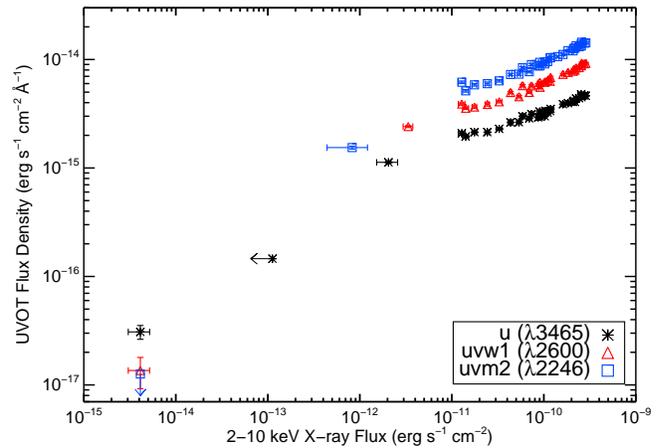}
 \caption{The UV/X-ray luminosity correlation for \src.  The UV filters (from Swift/UVOT) and their effective wavelengths are noted in the legend.  Hard state data is taken from \citet{armas-padilla13} during the 2011 outburst decay (we only include filters for which we have quiescent observations), and the data points at the lowest X-ray flux are from our recent campaign in quiescence.  The shape of the UV spectrum changes in quiescence and becomes redder (i.e., steeper).}
 \label{fig:luvlx}
 \end{figure}

\subsection{Multiwavelength Correlations}
\label{sec:res:corrs}
\subsubsection{Radio/X-ray}
\label{sec:res:radio}
In Figure~\ref{fig:lrlx} we show \src\ in the radio/X-ray luminosity plane, with luminosities illustrated from 2.3-6.3~kpc.   Even considering the distance uncertainty, our (3$\sigma$) radio limit on \src\ in quiescence (blue stars) indicates that its $\lr/\lx$ ratio is not any larger than  \srcasix.  Near the peak of the 2011 outburst,  a radio counterpart was detected from \src\  in the hard state on 4 Feb 2011, with a flux density of $245\pm54$ $\mu$Jy \citep{sivakoff11}.  We combine that radio flux density with a hard-state  \textit{XMM-Newton} observation taken  on 5 Feb 2011 \citep[red square;][unabsorbed $f_{\rm 0.5-10~keV}=3.3\times10^{-10}~\flux$]{armas-padilla14}.   Regardless of the distance, it is clear from Figure~\ref{fig:lrlx} that \src\ fell on the `radio-underluminous' branch of the $\lr-\lx$ diagram in the hard state.  Our new VLA observation in quiescence is therefore the most sensitive radio constraint on a  \xrb\ known to be radio-faint in the hard state although it is unclear if \src\ remained radio-underluminous in quiescence, or if it transitioned to the `traditional' radio/X-ray correlation.

\subsubsection{UV/X-ray}
\label{sec:res:uv}
 \src\ is only the second \xrb\  with quasi-simultaneous UV and X-ray detections deep in quiescence (after \srcxte; \citealt{mcclintock03, plotkin15}).  Only \src\  also has simultaneous high-cadence UV and X-ray monitoring observations during the outburst decay \citep{armas-padilla13}.     The decay UV/X-ray correlation from \citet{armas-padilla13} is shown in Figure~\ref{fig:luvlx}, along with our new data point in quiescence.  It is clear that the UV/X-ray correlation does not extend unbroken into quiescence.  Instead, the slope  steepens between the final hard state monitoring observations and our new quiescent epoch.  Interestingly, the shape of the UV spectrum also appears to have changed: in the hard state, the flux is brightest in the $uvm2$ filter, fainter in the $uvw1$ filter, and faintest in the $u$ filter; the opposite is observed in quiescence.

  \begin{figure}
\includegraphics[trim=1.5cm 1.0cm 0cm 0cm, clip=true, scale=0.5]{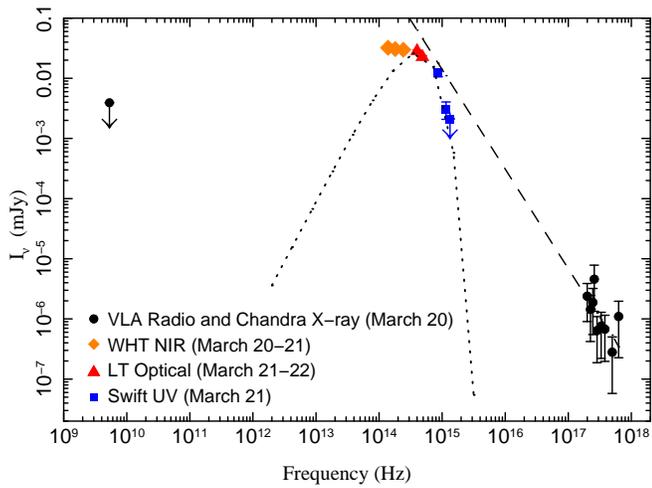} 
 \caption{Broadband quiescent spectrum of \src\ from our quasi-simultaneous multiwavelength campaign in March 2014. Data points are  coded by date of observation (see legend for details).  Chandra data are unfolded, binned to $>$2 counts per bin.   We also show our best-fit X-ray power-law spectrum ($\Gamma=2.6$; dashed line) extrapolated to lower frequencies, and a blackbody  curve ($T_{\rm bb}= 7000 \pm 300$~K) that fits the optical/UV data (dotted line).  We illustrate the X-ray spectrum with the best-fit $\Gamma=2.6$ instead of the canonical $\Gamma=2.1$ adopted elsewhere in the paper, in order to more conservatively illustrate that the UV spectrum is steeper than the X-ray spectrum.}
 \label{fig:srcsed}
 \end{figure}

\subsection{Broadband Spectrum}
\label{sec:res:SED}
The quiescent broadband spectrum is shown in Figure~\ref{fig:srcsed}, where it can be seen that the NIR, optical, and UV spectra become increasingly steeper.  The NIR spectrum is consistent with being flat (i.e., $f_\nu \propto$ constant): from our  $K_S$ and $J$-band observations, we measure a NIR spectral index $\alpha_{\rm nir}=  -0.1\pm0.1$ ($f_\nu \propto \nu^{\alpha}$).  Our optical $r^\prime$ and $i^\prime$ observations (which were taken on the same observing night) follow a steeper spectrum, $\alpha_{\rm opt}=-1.2\pm0.3$, which is consistent with the steep spectrum seen at earlier epochs (\citealt{shahbaz13} found $f_\nu \propto \nu^{-1.4}$).   Our   Swift/UVOT observations  indicate that the spectrum becomes even steeper at the highest observed UV energies ($\alpha_{\rm uv}<-2.6$ between the $uvw1$ and $uvm2$ filters).  

Given the change in spectral index between the NIR and UV bands, we attempt to fit the  NIR--UV spectrum with a single temperature  blackbody, but all  fits   are poor and underpedict the NIR flux.  We next explore the possibility that a single temperature blackbody explains only the optical-UV radiation, and we find  $T_{\rm bb}= 7000\pm300$ K from a least-squares fit \citep{markwardt09}.    Since there is likely day-long variability between the quasi-simultaneous optical and UV epochs, we do not expect to obtain a reduced $\chi^2_r \approx1$ during the fit (i.e., our statistical error bars  underestimate the uncertainty because they  neglect systematic errors; see, e.g.,  \citealt{markoff15}).  To derive the uncertainty on the best-fit $T_{\rm bb}$, we therefore adopt an empirical scheme where we fix the blackbody normalization to the best-fit value and  calculate blackbody curves for a grid of temperatures centered on 7000~K.  For the  99.7\% (i.e., 3$\sigma$) confidence interval, we adopt the range of temperatures that produce blackbody curves that pass through the $\pm 3\sigma$ error limits of at least one (detected) optical-UV data point, which corresponds to $6100 < T_{\rm bb} < 7900$~K (i.e., $\sigma_{T_{\rm bb}} = \pm300$ K).   In all cases, if the optical-UV emission were to be blackbody radiation from a cool accretion disk, then an extra emission component would be necessary to also explain the NIR spectrum (see Figure~\ref{fig:srcsed}).  We suggest in Section~\ref{sec:synchrotron} that synchrotron radiation can alternatively explain the entire NIR-UV spectrum, without any need to include additional components.

  \begin{figure}
\includegraphics[trim=1.5cm 1.0cm 0cm 0cm, clip=true, scale=0.5]{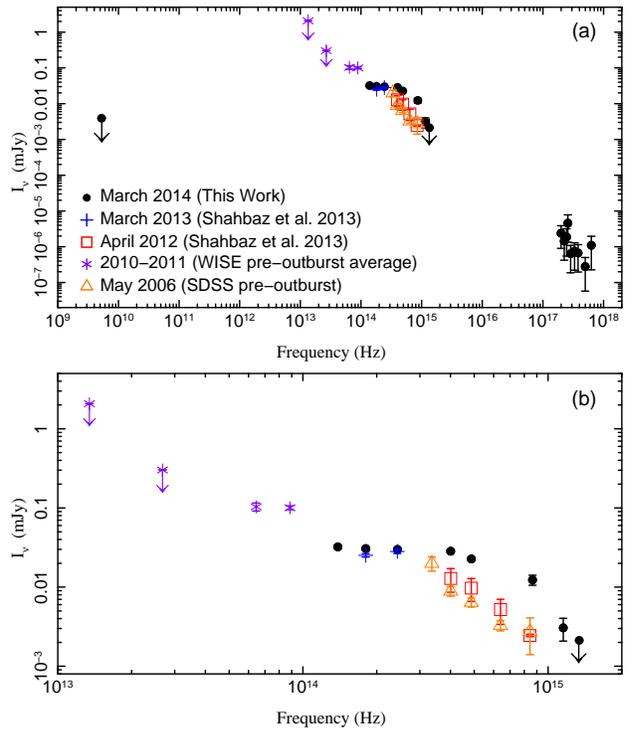}   
 \caption{Broadband spectrum of \src\ with archival IR, NIR, and optical observations.  Black circles are data from our 2014 quiescent campaign (repeated from Figure~\ref{fig:srcsed}).   Other symbols denote NIR and optical data at previous epochs (as published by  \citealt{shahbaz13}),   pre-outburst optical data from the SDSS, and pre-outburst IR observations from WISE (see legend for details).  For clarity, panel (b) shows a zoom-in of the IR-UV portion of the spectrum.}
 \label{fig:sed}
 \end{figure}

\begin{figure*}
\includegraphics[scale=0.65]{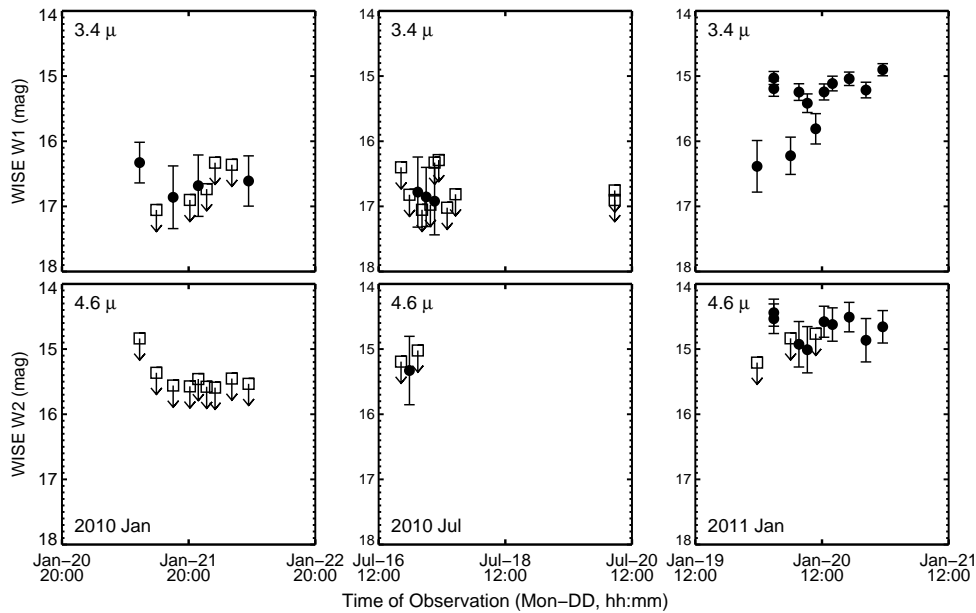}   
 \caption{Pre-outburst IR light curves of \src\ in the WISE W1 (top row) and W2 (bottom row) filters. Observations taken during 2010 January, 2010 July, and 2011 January are shown in the left, middle, and right columns, respectively.  Filled circles with error bars illustrate detections, while open squares represent 95\% lower limits on the magnitude.  The 2011 January observations were taken only eight days before \src\ was discovered in outburst in the X-ray, indicating that the outburst rise timescale was  $\lesssim$1 week.}
 \label{fig:wise}
 \end{figure*}

\subsection{Flux Variability}
\label{sec:res:var}
In Figure~\ref{fig:sed} we compare our 2014  broadband spectrum to previous IR through optical observations in quiescence.  These archival observations include pre-outburst data of the optical counterpart from the SDSS (2006), and post-outburst NIR and optical quiescent observations from 2012-2013 taken by \citet{shahbaz13}.  We only incorporate archival data with exposures $>$20 min in at least two filters on the same observing night, in order  to retain some simultaneous spectral information, and to average over short-term variability at a level comparable to our 2014 campaign (our shortest 2014 observation was 23 min).  We also include in Figure~\ref{fig:sed} (pre-outburst) IR observations from the Wide-field Infrared Survey Explorer \citep[WISE;][]{wright10}.

\src\  is a factor of 2-5 brighter at all frequencies during our 2014 campaign except for the NIR.   The \textit{Chandra} X-ray flux is almost a factor of three brighter than the quiescent flux from an \textit{XMM-Newton} observation obtained nine months  earlier (\citealt{armas-padilla14a} find $f_X = 3.2\times10^{-15}~\flux$, compared to $f_X= 8.2\times10^{-15}~\flux$ from \textit{Chandra}). This level of X-ray variability is typical for quiescent \xrb\ systems \citep[e.g.,][]{hynes04, bernardini14}.  On the same day as the \textit{XMM-Newton} observation, a fainter optical flux of $r^\prime=22.29\pm0.08$~mag (omitted from Figure~\ref{fig:sed} for clarity) was  observed with the LT \citep{armas-padilla14a};  our 2014 $r^\prime$ flux measurement is 4-5 times brighter.   Our other optical LT and  UVOT $u$ flux measurements are also  brighter by a similar factor.   Interestingly, the NIR data points do \textit{not} display the same level of long-term variability (and the 2014 NIR data also appears to follow a relatively flat spectrum).  However, no  epochs besides our 2014 campaign have coordinated NIR and optical data, so it is possible that the previous NIR observations were also taken  when the source was relatively bright.    The apparently elevated WISE IR flux densities are discussed in Section~\ref{sec:wisedisc}.  

\section{Archival Pre-Outburst Infrared Observations}
\label{sec:wisedisc}

 WISE  took IR observations on three separate epochs, covering 2010 January, 2010 July, and 2011 January (we take flux densities directly from the ALLWISE\footnote{\url{http://wise2.ipac.caltech.edu/docs/release/allwise/}} 
data release; we note that the WISE constraints on \src\ presented by \citealt{shahbaz13} only include data from the 2010 January epoch, which were the only publicly available data at the time).  Flux densities from co-added images of all three WISE epochs appear consistent with a flat IR spectrum (Figure~\ref{fig:sed}), albeit at a higher flux density compared  to the flat NIR spectrum that we observed in 2014.  

Serendipitously, the final WISE observations were taken on 2011 January 20, only eight days before \src\ was discovered in outburst in the X-ray.   We show IR light curves in the WISE W1 (3.4 $\mu$) and W2 (4.6$\mu$) filters in Figure~\ref{fig:wise} (the longer-wavelength WISE filter channels were not operational during the  2011 epoch), and  the 2011 IR flux density is clearly  elevated compared to the earlier 2010 WISE epochs.    Many of the WISE IR magnitudes are lower limits, so in order to quantify differences in IR fluxes between WISE epochs, we use the Astronomy SURVival Analysis (ASURV) package rev 1.2  \citep{lavalley92}, which implements the statistical methods presented in \citet{feigelson85}.  In the following, we  compare the 2011 January epoch (12  data points in each filter) to all 2010 epochs (22 data points in the W1 filter, and 12 data points in the W2 filter).\footnote{We combine both 2010 epochs here to improve the statistics, because we do not see  evidence that the distributions of WISE magnitudes are different between 2010 January and 2010 July.  A Peto-Prentence test indicates that the 2010 January and 2010 July magnitudes follow similar distributions ($p=0.1$ and $p=0.03$  for the WISE W1 and W2 filters, respectively).} 
    The IR magnitudes are  brighter during 2011:  incorporating the lower-limits, the mean magnitudes on 2011 January 20 are $\left<W_1\right>=15.40\pm0.13$ and $\left<W_2\right>=14.78\pm0.073$, compared to $\left<W_1\right>=16.90\pm0.047$ and $\left<W_2\right>=15.56\pm0.027$~mag during 2010.  Furthermore, a Peto-Prentence test indicates that the WISE magnitudes in 2011 vs.\ 2010 are unlikely drawn from the same parent distributions ($p<10^{-4}$ for both the W1 and W2 filters).
    
The elevated IR flux during the 2011 WISE epoch  might represent the early stages of the outburst.  However, since the 2011 IR light curve does not monotonically increase with time, we cannot exclude the possibility that the enhanced IR flux is a result of typical levels of variability expected in quiescence.  Regardless, the fact that \src\ was still so faint in the IR only eight days before the X-ray outburst discovery (which occurred when \src\ was already close to the peak of the X-ray outburst; \citealt{krimm11}) is consistent with expectations that \xrb\ outbursts rise quickly and decay slowly  \citep{chen97}. 

\section{Discussion}
\label{sec:disc}

We have presented a new, quasi-simultaneous  broadband spectrum covering the radio through X-ray wavebands for the \xrb\ \src\ in quiescence.    Our dataset includes the first UV constraints in quiescence, as well as the deepest radio observation to date (and to our knowledge, the deepest radio observation for any \xrb\ in quiescence that was known to be `radio-underluminous' in the hard state). The Eddington ratio during the observations was between   $10^{-8.4}\lesssim \lledd \lesssim 10^{-7.5}$, depending on the distance (2.3-6.3~kpc).   The optical through X-ray fluxes are a factor of 2-5 times brighter during our observing campaign compared to previous epochs.  However, no radio emission was detected in quiescence.    It is unlikely that  radio jet synchrotron emission is not detected simply because the proposed high-inclination nature of \src\ causes the jet to be beamed away from our line of sight.   \srcxte\ is constrained to have a similar inclination as \src\ ($68^\circ < i < 79^\circ$ for \srcxte; \citealt{khargharia13}), yet  radio emission is detected from \srcxte\ at $\lledd=10^{-8.5}$ \citep{gallo14}.  If \src\ launches a radio jet in quiescence, then the lack of a radio detection is either because \src\ is at too large of a distance to detect the radio jet, and/or  because of intrinsic factors (e.g., lower bulk jet speeds,  a jet axis that is not perpendicular to the orbital plane, etc., see, e.g., \citealt{gallo14}).  

\citet{armas-padilla13}  found that \src\ displayed a tight UV/X-ray luminosity correlation during its outburst decay (see Figure~\ref{fig:luvlx}), from which they determine that the UV radiation is dominated by a viscous accretion disk \citep[also see][]{weng15}.   Our observations show that the quiescent UV flux is almost two orders of magnitude fainter than the extrapolation of the UV/X-ray correlation to low X-ray fluxes, and the shape of the quiescent UV spectrum is also steeper.      An outer UV accretion disk is likely  present in quiescence (\citealt{hynes12}, also see, e.g.,  \citealt{mcclintock03, froning11}, who detect broad UV emission lines in the quiescent UV spectra of both \srcxte\ and \srcasix).   If  the outer accretion disk were to account for all of the  UV emission in quiescence, then the observed ``inversion'' in UV color between the hard state and quiescence would require a rapid decrease in the disk temperature, so that the UV probes the Rayleigh-Jeans tail in the hard state and the Wien tail in quiescence (the UV is very unlikely dominated by thermal emission from the companion star, given the companion's low stellar mass,   the lack of observed orbital NIR/optical flux modulations, and the lack of stellar absorption features in the quiescent optical spectrum).   While an outer disk almost certainly contributes some flux to the UV waveband, it cannot also explain the flat NIR spectrum in quiescence, and an extra emission component would be required. We argue below that, instead,  synchrotron radiation provides a natural explanation for the entire SED, including the inversion in UV color.

\subsection{A Synchrotron Origin for Quiescent \xrb\ Emission}
\label{sec:synchrotron}

 \citet{plotkin15} applied a broadband spectral model to \srcxte, which is the only other quiescent \xrb\ at an Eddington ratio comparable to \src\ that has a similarly rich, nearly simultaneous multiwavelength dataset.   \citet{plotkin15} conclude that the optical-UV spectrum of \srcxte\  arises from optically thin synchrotron radiation emitted by  a mildly relativistic population of thermal electrons.    These thermal electrons are  associated with the jet base in the analysis of \citet[][see \citealt{markoff05} for details]{plotkin15}.  They also allow for a weakly accelerated, non-thermal tail of particles in the outer jet (at gravitational radii $\rg \gtrsim 10-10^2$) that attain maximum electron Lorentz factors $\gamma_{e, \rm{max}} \lesssim 150$.  This particle acceleration is too weak, however, for the non-thermal tail to contribute significant amounts of synchrotron radiation to the X-ray band.  Therefore,  in the above scenario for \srcxte, the observed X-ray emission is  synchrotron self-Compton (SSC) emitted by the same thermal electrons responsible for the synchrotron optical-UV emission.  The same model applied to \srcasix\ suggests a similar geometry for that source in quiescence as well \citep[][]{gallo07}.

  We attempted to apply the same  broadband  model toward the SED of \src\  in quiescence.  However,  uncertainties on the distance, orbital inclination, and black hole mass leave too much degenerative parameter space to search through, preventing us from   obtaining a fit with meaningful constraints.   Instead, we phenomenologically interpret the broadband spectrum  below in order to highlight the most robust, model-independent results.  We defer detailed broadband modeling until the system parameters are better constrained.
  
  Our  conclusion for synchrotron-dominated optical/UV radiation that is emitted by a thermal population of electrons agrees with the  conclusions of \citet{shahbaz13}.  \citet{shahbaz13} argue from a steep spectrum and short-term variability that synchrotron emission from thermal electrons dominates the  optical emission from \src\ in quiescence.  They similarly associate this thermal synchrotron emission with an outflowing  jet  instead of an inflowing RIAF:   synchrotron radiation from the simplest class of RIAFs  (i.e., the advection dominated accretion flow; ADAF; e.g., \citealt{narayan94}) is expected to display a sharp peak due to synchrotron self-absorption  near $\nu_p \sim 10^{15} \left(M/\msun\right)^{-1/2}$~Hz, which is not observed in the broadband spectrum.   We note that \citet{shahbaz13} analyzed a non-simultaneous multiwavelength dataset that did not contain radio, UV, and X-ray constraints.  It is therefore possible for us to obtain a more complete picture of the quiescent jet from our quasi-simultaneous broadband spectrum.

 We  expect the synchrotron jet emission to  become optically thick at low frequencies.  In other words, provided that multiple `zones'  along the jet contribute synchrotron radiation (with the peak temperature of the thermal electrons decreasing for zones farther from the black hole), the superposition of multiple synchrotron self-absorbed spectra will eventually create a flat spectrum at frequencies below a break frequency $\nu_b$ (where $\nu_b$ marks the optically thick transition; e.g., \citealt{blandford79}).   Extrapolating  the NIR and optical spectra  ($\alpha_{\rm nir}=-0.1\pm0.1$ and $\alpha_{\rm nir}=-1.2\pm0.3$; see Section~\ref{sec:res:SED}) to higher and lower frequencies, respectively, we find that $\nu_b \approx \left(4.1 \pm 0.6\right) \times10^{14}$~Hz.   This calculation of $\nu_b$ neglects systematics due to variability, and the quoted (statistical) error is likely underestimated.  Visual inspection of Figure~\ref{fig:sed}b indicates that $\nu_b$ must occur between the NIR $J$ and optical $r^\prime$ filters.  We therefore adopt a more conservative estimate of $\nu_b$ to be between $\approx2-5 \times 10^{14}$ Hz.   We note that our radio limit constrains  the outer jet to follow an optically thick, inverted spectrum with $\alpha_\nu > +0.3$ ($f_\nu \propto \nu^{\alpha_\nu}$) between the radio and IR/NIR wavebands, which further supports the idea of optically thick synchrotron radiation at low frequencies.  
 
 \citet{shahbaz13} concluded that the jet break should occur at $\nu_b < 2.5\times10^{13}$~Hz in their SED (at least an order of magnitude lower frequency than our constraint).  Our 2014 jet break detection may therefore imply that $\nu_b$  varies with time, shifting through the IR and NIR bands.  Rapid variability of $\nu_b$ through the IR has also been observed for GX 339-4 in the hard state, which is most likely caused by fast variations in the strength of the jet magnetic field \citep{gandhi11}.  We note that the \citet{shahbaz13} NIR data points from March 2013 also appear to follow a flat spectrum in Figure~\ref{fig:sed}b, which  further supports a variable jet break frequency (the rest of the multiwavelength data that their $\nu_b<2.5\times10^{13}$~Hz limit is based on was obtained 1-3 years prior to their 2013 NIR observations).
   
If the approximately flat NIR spectrum is  due to an optically thick jet\footnote{It is also a possibility that the flat NIR spectrum could rather  be associated with an entirely optically thin synchrotron  component (emitted by the thermal jet) that simply peaks near the NIR.}, then this would mark the first time that an optically thick-to-thin jet break has been directly observed in quiescence (emission from the companion stars in \srcasix\ and \srcxte\ prohibit jet break identifications in those systems without broadband spectral modeling).   Jet breaks have been isolated in almost a dozen hard state systems \citep[e.g.,][]{corbel02, russell13}.  However, we caution that drawing a physical connection between a  jet break for \src\ in quiescence and for  the hard state systems is currently premature, as the jet breaks that have been isolated in hard state systems are intimately related to the location along the jet where (some fraction of) particles are accelerated into non-thermal distributions \citep[e.g.,][]{polko13, polko14}.  For \src\ it is unclear if such an acceleration zone exists in quiescence, and the potential jet break may rather simply represent the optically thick to thin transition of a multi-zone jet composed purely of a thermal distribution of particles. 

\subsubsection{X-ray Radiation Mechanisms in Quiescence}
\label{sec:disc:xray}
Interpretation of the X-ray emission highly depends on the efficiency of  particle acceleration along the jet (i.e., the maximum Lorentz factor, $\gamma_{e,{\rm max}}$ attained by  accelerated particles).   If particle acceleration is weak  (with $\gamma_{e,{\rm max}}<10^2$, as suggested by \citealt{plotkin15} for \srcxte), then (non-thermal) synchrotron radiation from the jet will quickly cool  and not contribute significantly at X-ray energies.  The optical-UV spectrum would be synchrotron radiation from thermal electrons as described above, and the X-rays must be  corresponding (thermal) SSC. 

If particle acceleration is instead efficient (i.e., $\gamma_{e,{\rm max}}>>10^{2}$), then synchrotron emission  from non-thermal electrons (and/or corresponding SSC) along the accelerated jet would be responsible for most of the observed  X-rays.  However, the (steep) \textit{Chandra} X-ray spectrum  implies that any non-thermal electrons must suffer from radiative  losses, and the X-rays would be synchrotron cooled.\footnote{Electrons injected into a  non-thermal distribution by, e.g., diffusive shock acceleration processes will follow a particle spectrum $n_e\left(\gamma_e\right) \propto \gamma_e^{-p}$, where $n_e$ is the electron density, $\gamma_e$ is the Lorentz factor of each electron, and the spectral slope $p$ is typically 2-2.4.  Radiative cooling losses will modify the spectral slope to steepen by $p+1$ at X-ray energies, yielding an X-ray photon index for synchrotron cooled jet emission of $\Gamma=\left(p+2\right)/2 \approx 2.$} 
In that case, the slope of the radio/X-ray correlation would steepen by a factor of $\approx$2 in quiescence \citep{yuan05a}, and the corresponding optically thick synchrotron  radio emission from the outer jet would be well below the sensitivity of our VLA observations.   

Either efficient or inefficient particle acceleration along the jet is consistent with the observed X-ray spectrum.  However, in both scenarios, the NIR-UV emission must always be dominated by synchrotron radiation from  \textit{thermal} electrons.  To illustrate this point, it is clear from Figure~\ref{fig:srcsed} that the UV emission follows a steeper spectral slope than the X-ray emission.   The UV emission therefore cannot also be synchrotron cooled radiation from the same non-thermal electrons (or else the UV and X-ray radiation would follow similar spectral slopes).    In order for X-rays to be emitted by synchrotron cooled radiation,   the synchrotron cooling break must fall below the X-ray waveband (i.e., $\nu_{\rm cool} \lesssim 10^{17}$~Hz).  At frequencies below the cooling break, synchrotron radiation from accelerated non-thermal electrons would be optically thin (i.e., the spectrum would become \textit{flatter}, with a photon index smaller by $\Delta \Gamma=0.5$; e.g., \citealt{rybicki79}).   To explain the observed shape (and especially curvature) of the optical/UV spectrum,  the thermal synchrotron component must dominate over any optically thin synchrotron radiation emitted by the accelerated particles.  \textit{Thus, regardless of the particle acceleration efficiency and the mechanism that produces the X-ray radiation,   emission from a thermal (relativistic) population of electrons always  dominates the lower-energy radiation. }

\section{Conclusions}
\label{sec:conc}
From a new quasi-simultaneous SED of \src\ in quiescence, we obtain the following:

\begin{itemize}

\item We isolate the frequency where the jet transitions from optically thick-to-thin to be   $\nu_b \approx 2-5 \times10^{14}$~Hz, which represents the first direct detection of a jet break in a  quiescent \xrb.  Comparing to the  $\nu_b < 2.5 \times 10^{13}$~Hz limit placed by \citet{shahbaz13} at earlier epochs, our detection suggests a variable jet break frequency in quiescence that shifts through the IR/NIR bands.

\item Nearly simultaneous UV and X-ray observations show a switch in the UV radiation mechanism in quiescence compared to the hard state.  This result is consistent with  a thermal synchrotron ``bump'' peaking near the optical (akin to the ``sub-mm bump'' in Sgr A*; e.g., \citealt{serabyn97, falcke00}), which  could be a common feature of quiescent, short-period low-mass \xrb s (such a feature is also seen for \srcxte\ and \srcasix\ in quiescence).   We demonstrate that the thermal synchrotron origin of this component does not depend on the  details of the X-ray emission mechanism.

\item We provided the deepest radio limit yet for any \xrb\ in quiescence that was known to be ``radio-underluminous'' in the hard state.  This limit indicates that the outer (quiescent) jet   follows an inverted, optically-thick synchrotron spectrum  ($\alpha_{\nu} > 0.3$).  

\end{itemize}

Finally, we conclude by noting that our UV constraints were crucial for reaching many of our conclusions  (especially the second bullet point above).  Without the UV constraints on the curvature of the optical--UV spectrum, it would be tempting to connect the optical and X-ray emission with a simple (synchrotron cooled) power-law, while attributing the difference between optical and X-ray flux normalizations to day-long variability between the two wavebands. High Galactic latitude sources are therefore critical for opening the UV window and understanding very low Eddington ratio accretion flows \citep[also see][]{mcclintock03}.  This key point should serve as further motivation for black hole surveys targeting high Galactic latitudes.  

\section*{Acknowledgments}
We thank the anonymous referee for insightful comments that improved this manuscript. We greatly appreciate the efforts of the CXC and NRAO schedulers for coordinating the simultaneous \textit{Chandra} and VLA observations.  We also thank Neil Gehrels and the Swift team for approving and scheduling the Swift/UVOT observations.  We thank Adam Kowalski for helpful discussions regarding   M-stars, and Edmund Hodges-Kluck for advice related to Swift/UVOT.  Support for this work was provided by the National Aeronautics and Space Administration through Chandra Award Number GO4-15042X issued by the Chandra X-ray Observatory Center, which is operated by the Smithsonian Astrophysical Observatory for and on behalf of the National Aeronautics Space Administration under contract NAS8-03060.  JCAMJ is the recipient of an Australian Research Council Future Fellowship (FT140101082).  TMD acknowledges support by the Spanish Ministerio de Econom'a y competitividad (MINECO) under grant AYA2013-42627.  This research has made use of software provided by the Chandra X-ray Center (CXC) in the application package CIAO.  The William Herschel Telescope is operated on the island of La Palma by the Isaac Newton Group in the Spanish Observatorio del Roque de los Muchachos of the Instituto de Astrof'sica de Canarias.  The National Radio Astronomy Observatory is a facility of the National Science Foundation operated under cooperative agreement by Associated Universities, Inc.  This publication makes use of data products from the Wide-field Infrared Survey Explorer, which is a joint project of the University of California, Los Angeles, and the Jet Propulsion Laboratory/California Institute of Technology, funded by the National Aeronautics and Space Administration.

\bibliographystyle{mn2e}

\begin{thebibliography}{90}
\expandafter\ifx\csname natexlab\endcsname\relax\def\natexlab#1{#1}\fi

\bibitem[{{Abramowicz} {et~al}\mbox{.}(1995){Abramowicz}, {Chen}, {Kato},
  {Lasota}, \& {Regev}}]{abramowicz95}
{Abramowicz} M.~A., {Chen} X., {Kato} S., {Lasota} J.-P., {Regev} O., 1995,
  \apjl, 438, L37

\bibitem[{{Armas Padilla} {et~al}\mbox{.}(2013){Armas Padilla}, {Degenaar},
  {Russell}, \& {Wijnands}}]{armas-padilla13}
{Armas Padilla} M., {Degenaar} N., {Russell} D.~M., {Wijnands} R., 2013,
  \mnras, 428, 3083

\bibitem[{{Armas Padilla} {et~al}\mbox{.}(2014{\natexlab{a}}){Armas Padilla},
  {Wijnands}, {Altamirano}, {M{\'e}ndez}, {Miller}, \&
  {Degenaar}}]{armas-padilla14}
{Armas Padilla} M., {Wijnands} R., {Altamirano} D., {M{\'e}ndez} M., {Miller}
  J.~M., {Degenaar} N., 2014{\natexlab{a}}, \mnras, 439, 3908

\bibitem[{{Armas Padilla} {et~al}\mbox{.}(2014{\natexlab{b}}){Armas Padilla},
  {Wijnands}, {Degenaar}, {Mu{\~n}oz-Darias}, {Casares}, \&
  {Fender}}]{armas-padilla14a}
{Armas Padilla} M., {Wijnands} R., {Degenaar} N., {Mu{\~n}oz-Darias} T.,
  {Casares} J., {Fender} R.~P., 2014{\natexlab{b}}, \mnras, 444, 902

\bibitem[{{Bernardini} \& {Cackett}(2014)}]{bernardini14}
{Bernardini} F., {Cackett} E.~M., 2014, \mnras, 439, 2771

\bibitem[{{Blandford} \& {Begelman}(1999)}]{blandford99}
{Blandford} R.~D., {Begelman} M.~C., 1999, \mnras, 303, L1

\bibitem[{{Blandford} \& {K{\"o}nigl}(1979)}]{blandford79}
{Blandford} R.~D., {K{\"o}nigl} A., 1979, \apj, 232, 34

\bibitem[{{Brocksopp} {et~al}\mbox{.}(2005){Brocksopp}, {Corbel}, {Fender},
  {Rupen}, {Sault}, {Tingay}, {Hannikainen}, \& {O'Brien}}]{brocksopp05}
{Brocksopp} C., {Corbel} S., {Fender} R.~P., {Rupen} M., {Sault} R., {Tingay}
  S.~J., {Hannikainen} D., {O'Brien} K., 2005, \mnras, 356, 125

\bibitem[{{Cadolle Bel} {et~al}\mbox{.}(2007){Cadolle Bel}, {Rib{\'o}},
  {Rodriguez}, {Chaty}, {Corbel}, {Goldwurm}, {Frontera}, {Farinelli},
  {D'Avanzo}, {Tarana}, {Ubertini}, {Laurent}, {Goldoni}, \&
  {Mirabel}}]{cadolle-bel07}
{Cadolle Bel} M., {Rib{\'o}} M., {Rodriguez} J. {et~al.}, 2007, \apj, 659, 549

\bibitem[{{Calvelo} {et~al}\mbox{.}(2010){Calvelo}, {Fender}, {Russell},
  {Gallo}, {Corbel}, {Tzioumis}, {Bell}, {Lewis}, \& {Maccarone}}]{calvelo10}
{Calvelo} D.~E., {Fender} R.~P., {Russell} D.~M. {et~al.}, 2010, \mnras, 409,
  839

\bibitem[{{Cao}, {Wu} \& {Dong}(2014){Cao}, {Wu}, \& {Dong}}]{cao14}
{Cao} X.-F., {Wu} Q., {Dong} A.-J., 2014, \apj, 788, 52

\bibitem[{{Cash}(1979)}]{cash79}
{Cash} W., 1979, \apj, 228, 939

\bibitem[{{Chen}, {Shrader} \& {Livio}(1997){Chen}, {Shrader}, \&
  {Livio}}]{chen97}
{Chen} W., {Shrader} C.~R., {Livio} M., 1997, \apj, 491, 312

\bibitem[{{Corbel} {et~al}\mbox{.}(2013){Corbel}, {Coriat}, {Brocksopp},
  {Tzioumis}, {Fender}, {Tomsick}, {Buxton}, \& {Bailyn}}]{corbel13}
{Corbel} S., {Coriat} M., {Brocksopp} C., {Tzioumis} A.~K., {Fender} R.~P.,
  {Tomsick} J.~A., {Buxton} M.~M., {Bailyn} C.~D., 2013, \mnras, 428, 2500

\bibitem[{{Corbel} {et~al}\mbox{.}(2004){Corbel}, {Fender}, {Tomsick},
  {Tzioumis}, \& {Tingay}}]{corbel04}
{Corbel} S., {Fender} R.~P., {Tomsick} J.~A., {Tzioumis} A.~K., {Tingay} S.,
  2004, \apj, 617, 1272

\bibitem[{{Corbel} {et~al}\mbox{.}(2002){Corbel}, {Fender}, {Tzioumis},
  {Tomsick}, {Orosz}, {Miller}, {Wijnands}, \& {Kaaret}}]{corbel02}
{Corbel} S., {Fender} R.~P., {Tzioumis} A.~K., {Tomsick} J.~A., {Orosz} J.~A.,
  {Miller} J.~M., {Wijnands} R., {Kaaret} P., 2002, Science, 298, 196

\bibitem[{{Coriat} {et~al}\mbox{.}(2011){Coriat}, {Corbel}, {Prat},
  {Miller-Jones}, {Cseh}, {Tzioumis}, {Brocksopp}, {Rodriguez}, {Fender}, \&
  {Sivakoff}}]{coriat11}
{Coriat} M., {Corbel} S., {Prat} L. {et~al.}, 2011, \mnras, 414, 677

\bibitem[{{Corral-Santana} {et~al}\mbox{.}(2013){Corral-Santana}, {Casares},
  {Mu{\~n}oz-Darias}, {Rodr{\'{\i}}guez-Gil}, {Shahbaz}, {Torres}, {Zurita}, \&
  {Tyndall}}]{corral-santana13}
{Corral-Santana} J.~M., {Casares} J., {Mu{\~n}oz-Darias} T.,
  {Rodr{\'{\i}}guez-Gil} P., {Shahbaz} T., {Torres} M.~A.~P., {Zurita} C.,
  {Tyndall} A.~A., 2013, Science, 339, 1048

\bibitem[{{Dzib}, {Massi} \& {Jaron}(2015){Dzib}, {Massi}, \& {Jaron}}]{dzib15}
{Dzib} S.~A., {Massi} M., {Jaron} F., 2015, \aap, 580, L6

\bibitem[Falcke 
\& Markoff(2000)]{falcke00} Falcke, H., \& Markoff, S.\ 2000, \aap, 362, 113 

\bibitem[{{Feigelson} \& {Nelson}(1985)}]{feigelson85}
{Feigelson} E.~D., {Nelson} P.~I., 1985, \apj, 293, 192

\bibitem[{{Fender}(2001)}]{fender01}
{Fender} R.~P., 2001, \mnras, 322, 31

\bibitem[{{Froning} {et~al}\mbox{.}(2011){Froning}, {Cantrell}, {Maccarone},
  {France}, {Khargharia}, {Winter}, {Robinson}, {Hynes}, {Broderick},
  {Markoff}, {Torres}, {Garcia}, {Bailyn}, {Prochaska}, {Werk}, {Thom},
  {B{\'e}land}, {Danforth}, {Keeney}, \& {Green}}]{froning11}
{Froning} C.~S., {Cantrell} A.~G., {Maccarone} T.~J. {et~al.}, 2011, \apj, 743,
  26

\bibitem[{{Fruscione} {et~al}\mbox{.}(2006){Fruscione}, {McDowell}, {Allen},
  {Brickhouse}, {Burke}, {Davis}, {Durham}, {Elvis}, {Galle}, {Harris},
  {Huenemoerder}, {Houck}, {Ishibashi}, {Karovska}, {Nicastro}, {Noble},
  {Nowak}, {Primini}, {Siemiginowska}, {Smith}, \& {Wise}}]{fruscione06}
{Fruscione} A., {McDowell} J.~C., {Allen} G.~E. {et~al.}, 2006, in Society of
  Photo-Optical Instrumentation Engineers (SPIE) Conference Series, Vol. 6270,
  Society of Photo-Optical Instrumentation Engineers (SPIE) Conference Series,
  p.~1

\bibitem[{{Gallo}, {Fender} \& {Corbel}(2003){Gallo}, {Fender}, \&
  {Corbel}}]{gallo03}
{Gallo} E., {Fender} R., {Corbel} S., 2003, The Astronomer's Telegram, 196, 1

\bibitem[{{Gallo}, {Fender} \& {Hynes}(2005){Gallo}, {Fender}, \&
  {Hynes}}]{gallo05}
{Gallo} E., {Fender} R.~P., {Hynes} R.~I., 2005, \mnras, 356, 1017

\bibitem[{{Gallo} {et~al}\mbox{.}(2006){Gallo}, {Fender}, {Miller-Jones},
  {Merloni}, {Jonker}, {Heinz}, {Maccarone}, \& {van der Klis}}]{gallo06}
{Gallo} E., {Fender} R.~P., {Miller-Jones} J.~C.~A., {Merloni} A., {Jonker}
  P.~G., {Heinz} S., {Maccarone} T.~J., {van der Klis} M., 2006, \mnras, 370,
  1351

\bibitem[{{Gallo} {et~al}\mbox{.}(2007){Gallo}, {Migliari}, {Markoff},
  {Tomsick}, {Bailyn}, {Berta}, {Fender}, \& {Miller-Jones}}]{gallo07}
{Gallo} E., {Migliari} S., {Markoff} S., {Tomsick} J.~A., {Bailyn} C.~D.,
  {Berta} S., {Fender} R., {Miller-Jones} J.~C.~A., 2007, \apj, 670, 600

\bibitem[{{Gallo}, {Miller} \& {Fender}(2012){Gallo}, {Miller}, \&
  {Fender}}]{gallo12}
{Gallo} E., {Miller} B.~P., {Fender} R., 2012, \mnras, 423, 590

\bibitem[{{Gallo} {et~al}\mbox{.}(2014){Gallo}, {Miller-Jones}, {Russell},
  {Jonker}, {Homan}, {Plotkin}, {Markoff}, {Miller}, {Corbel}, \&
  {Fender}}]{gallo14}
{Gallo} E., {Miller-Jones} J.~C.~A., {Russell} D.~M. {et~al.}, 2014, \mnras,
  445, 290

\bibitem[{{Gandhi} {et~al}\mbox{.}(2011){Gandhi}, {Blain}, {Russell},
  {Casella}, {Malzac}, {Corbel}, {D'Avanzo}, {Lewis}, {Markoff}, {Cadolle Bel},
  {Goldoni}, {Wachter}, {Khangulyan}, \& {Mainzer}}]{gandhi11}
{Gandhi} P., {Blain} A.~W., {Russell} D.~M. {et~al.}, 2011, \apjl, 740, L13

\bibitem[{{Garmire} {et~al}\mbox{.}(2003){Garmire}, {Bautz}, {Ford}, {Nousek},
  \& {Ricker}}]{garmire03}
{Garmire} G.~P., {Bautz} M.~W., {Ford} P.~G., {Nousek} J.~A., {Ricker}, Jr.
  G.~R., 2003, in Society of Photo-Optical Instrumentation Engineers (SPIE)
  Conference Series, Vol. 4851, X-Ray and Gamma-Ray Telescopes and Instruments
  for Astronomy., {Truemper} J.~E., {Tananbaum} H.~D., eds., pp. 28--44

\bibitem[{{Heinz} \& {Sunyaev}(2003)}]{heinz03}
{Heinz} S., {Sunyaev} R.~A., 2003, \mnras, 343, L59

\bibitem[{{Hjellming} \& {Johnston}(1988)}]{hjellming88}
{Hjellming} R.~M., {Johnston} K.~J., 1988, \apj, 328, 600

\bibitem[{{Hjellming} {et~al}\mbox{.}(2000){Hjellming}, {Rupen},
  {Mioduszewski}, \& {Narayan}}]{hjellming00}
{Hjellming} R.~M., {Rupen} M.~P., {Mioduszewski} A.~J., {Narayan} R., 2000, The
  Astronomer's Telegram, 54, 1

\bibitem[{{Houck} \& {Denicola}(2000)}]{houck00}
{Houck} J.~C., {Denicola} L.~A., 2000, in Astronomical Society of the Pacific
  Conference Series, Vol. 216, Astronomical Data Analysis Software and Systems
  IX, {Manset} N., {Veillet} C., {Crabtree} D., eds., p. 591

\bibitem[{{Hynes} {et~al}\mbox{.}(2009){Hynes}, {Bradley}, {Rupen}, {Gallo},
  {Fender}, {Casares}, \& {Zurita}}]{hynes09}
{Hynes} R.~I., {Bradley} C.~K., {Rupen} M., {Gallo} E., {Fender} R.~P.,
  {Casares} J., {Zurita} C., 2009, \mnras, 399, 2239

\bibitem[{{Hynes} {et~al}\mbox{.}(2004){Hynes}, {Charles}, {Garcia},
  {Robinson}, {Casares}, {Haswell}, {Kong}, {Rupen}, {Fender}, {Wagner},
  {Gallo}, {Eves}, {Shahbaz}, \& {Zurita}}]{hynes04}
{Hynes} R.~I., {Charles} P.~A., {Garcia} M.~R. {et~al.}, 2004, \apjl, 611, L125

\bibitem[{{Hynes} \& {Robinson}(2012)}]{hynes12}
{Hynes} R.~I., {Robinson} E.~L., 2012, \apj, 749, 3

\bibitem[{{Ichimaru}(1977)}]{ichimaru77}
{Ichimaru} S., 1977, \apj, 214, 840

\bibitem[{{Jonker} {et~al}\mbox{.}(2004){Jonker}, {Gallo}, {Dhawan}, {Rupen},
  {Fender}, \& {Dubus}}]{jonker04}
{Jonker} P.~G., {Gallo} E., {Dhawan} V., {Rupen} M., {Fender} R.~P., {Dubus}
  G., 2004, \mnras, 351, 1359

\bibitem[{{Jonker} {et~al}\mbox{.}(2010){Jonker}, {Miller-Jones}, {Homan},
  {Gallo}, {Rupen}, {Tomsick}, {Fender}, {Kaaret}, {Steeghs}, {Torres},
  {Wijnands}, {Markoff}, \& {Lewin}}]{jonker10}
{Jonker} P.~G., {Miller-Jones} J., {Homan} J. {et~al.}, 2010, \mnras, 401, 1255

\bibitem[{{Jonker} {et~al}\mbox{.}(2012){Jonker}, {Miller-Jones}, {Homan},
  {Tomsick}, {Fender}, {Kaaret}, {Markoff}, \& {Gallo}}]{jonker12}
{Jonker} P.~G., {Miller-Jones} J.~C.~A., {Homan} J., {Tomsick} J., {Fender}
  R.~P., {Kaaret} P., {Markoff} S., {Gallo} E., 2012, \mnras, 423, 3308

\bibitem[{{Kataoka} {et~al}\mbox{.}(2008){Kataoka}, {Madejski}, {Sikora},
  {Roming}, {Chester}, {Grupe}, {Tsubuku}, {Sato}, {Kawai}, {Tosti},
  {Impiombato}, {Kovalev}, {Kovalev}, {Edwards}, {Wagner}, {Moderski},
  {Stawarz}, {Takahashi}, \& {Watanabe}}]{kataoka08}
{Kataoka} J., {Madejski} G., {Sikora} M. {et~al.}, 2008, \apj, 672, 787

\bibitem[{{Khargharia} {et~al}\mbox{.}(2013){Khargharia}, {Froning},
  {Robinson}, \& {Gelino}}]{khargharia13}
{Khargharia} J., {Froning} C.~S., {Robinson} E.~L., {Gelino} D.~M., 2013, \aj,
  145, 21

\bibitem[{{Krimm} {et~al}\mbox{.}(2011){Krimm}, {Barthelmy}, {Baumgartner},
  {Cummings}, {Fenimore}, {Gehrels}, {Markwardt}, {Palmer}, {Sakamoto},
  {Skinner}, {Stamatikos}, {Tueller}, \& {Ukwatta}}]{krimm11}
{Krimm} H.~A., {Barthelmy} S.~D., {Baumgartner} W. {et~al.}, 2011, The
  Astronomer's Telegram, 3138, 1

\bibitem[{{Krimm}, {Kennea} \& {Holland}(2011){Krimm}, {Kennea}, \&
  {Holland}}]{krimm11a}
{Krimm} H.~A., {Kennea} J.~A., {Holland} S.~T., 2011, The Astronomer's
  Telegram, 3142, 1

\bibitem[{{Lavalley}, {Isobe} \& {Feigelson}(1992){Lavalley}, {Isobe}, \&
  {Feigelson}}]{lavalley92}
{Lavalley} M., {Isobe} T., {Feigelson} E., 1992, in Astronomical Society of the
  Pacific Conference Series, Vol.~25, Astronomical Data Analysis Software and
  Systems I, {Worrall} D.~M., {Biemesderfer} C., {Barnes} J., eds., p. 245

\bibitem[{{Markoff}, {Falcke} \& {Fender}(2001){Markoff}, {Falcke}, \&
  {Fender}}]{markoff01}
{Markoff} S., {Falcke} H., {Fender} R., 2001, \aap, 372, L25

\bibitem[{{Markoff} {et~al}\mbox{.}(2003){Markoff}, {Nowak}, {Corbel},
  {Fender}, \& {Falcke}}]{markoff03}
{Markoff} S., {Nowak} M., {Corbel} S., {Fender} R., {Falcke} H., 2003, \aap,
  397, 645

\bibitem[{{Markoff} {et~al}\mbox{.}(2015){Markoff}, {Nowak}, {Gallo}, {Hynes},
  {Wilms}, {Plotkin}, {Maitra}, {Silva}, \& {Drappeau}}]{markoff15}
{Markoff} S., {Nowak} M.~A., {Gallo} E. {et~al.}, 2015, \apjl, 812, L25

\bibitem[{{Markoff}, {Nowak} \& {Wilms}(2005){Markoff}, {Nowak}, \&
  {Wilms}}]{markoff05}
{Markoff} S., {Nowak} M.~A., {Wilms} J., 2005, \apj, 635, 1203

\bibitem[{{Markwardt}(2009)}]{markwardt09}
{Markwardt} C.~B., 2009, in Astronomical Society of the Pacific Conference
  Series, Vol. 411, Astronomical Data Analysis Software and Systems XVIII,
  {Bohlender} D.~A., {Durand} D., {Dowler} P., eds., p. 251

\bibitem[{{Mata S{\'a}nchez} {et~al}\mbox{.}(2015){Mata S{\'a}nchez},
  {Mu{\~n}oz-Darias}, {Casares}, {Corral-Santana}, \&
  {Shahbaz}}]{mata-sanchez15}
{Mata S{\'a}nchez} D., {Mu{\~n}oz-Darias} T., {Casares} J., {Corral-Santana}
  J.~M., {Shahbaz} T., 2015, \mnras, 454, 2199

\bibitem[{{McClintock} {et~al}\mbox{.}(2003){McClintock}, {Narayan}, {Garcia},
  {Orosz}, {Remillard}, \& {Murray}}]{mcclintock03}
{McClintock} J.~E., {Narayan} R., {Garcia} M.~R., {Orosz} J.~A., {Remillard}
  R.~A., {Murray} S.~S., 2003, \apj, 593, 435

\bibitem[{{McMullin} {et~al}\mbox{.}(2007){McMullin}, {Waters}, {Schiebel},
  {Young}, \& {Golap}}]{mcmullin07}
{McMullin} J.~P., {Waters} B., {Schiebel} D., {Young} W., {Golap} K., 2007, in
  Astronomical Society of the Pacific Conference Series, Vol. 376, Astronomical
  Data Analysis Software and Systems XVI, {Shaw} R.~A., {Hill} F., {Bell}
  D.~J., eds., p. 127

\bibitem[{{Meyer-Hofmeister} \& {Meyer}(2014)}]{meyer-hofmeister14}
{Meyer-Hofmeister} E., {Meyer} F., 2014, \aap, 562, A142

\bibitem[{{Miller} {et~al}\mbox{.}(2006){Miller}, {Raymond}, {Homan}, {Fabian},
  {Steeghs}, {Wijnands}, {Rupen}, {Charles}, {van der Klis}, \&
  {Lewin}}]{miller06}
{Miller} J.~M., {Raymond} J., {Homan} J. {et~al.}, 2006, \apj, 646, 394

\bibitem[{{Miller-Jones} {et~al}\mbox{.}(2011){Miller-Jones}, {Jonker},
  {Maccarone}, {Nelemans}, \& {Calvelo}}]{miller-jones11}
{Miller-Jones} J.~C.~A., {Jonker} P.~G., {Maccarone} T.~J., {Nelemans} G.,
  {Calvelo} D.~E., 2011, \apjl, 739, L18

\bibitem[{{Munar-Adrover} {et~al}\mbox{.}(2014){Munar-Adrover}, {Paredes},
  {Rib{\'o}}, {Iwasawa}, {Zabalza}, \& {Casares}}]{munar-adrover14}
{Munar-Adrover} P., {Paredes} J.~M., {Rib{\'o}} M., {Iwasawa} K., {Zabalza} V.,
  {Casares} J., 2014, \apjl, 786, L11

\bibitem[{{Narayan}, {Igumenshchev} \& {Abramowicz}(2000){Narayan},
  {Igumenshchev}, \& {Abramowicz}}]{narayan00}
{Narayan} R., {Igumenshchev} I.~V., {Abramowicz} M.~A., 2000, \apj, 539, 798

\bibitem[{{Narayan} \& {Yi}(1994)}]{narayan94}
{Narayan} R., {Yi} I., 1994, \apjl, 428, L13

\bibitem[{{Perley} \& {Butler}(2013)}]{perley13}
{Perley} R.~A., {Butler} B.~J., 2013, \apjs, 204, 19

\bibitem[{{Plotkin}, {Gallo} \& {Jonker}(2013){Plotkin}, {Gallo}, \&
  {Jonker}}]{plotkin13}
{Plotkin} R.~M., {Gallo} E., {Jonker} P.~G., 2013, \apj, 773, 59

\bibitem[{{Plotkin} {et~al}\mbox{.}(2015){Plotkin}, {Gallo}, {Markoff},
  {Homan}, {Jonker}, {Miller-Jones}, {Russell}, \& {Drappeau}}]{plotkin15}
{Plotkin} R.~M., {Gallo} E., {Markoff} S., {Homan} J., {Jonker} P.~G.,
  {Miller-Jones} J.~C.~A., {Russell} D.~M., {Drappeau} S., 2015, \mnras, 446,
  4098

\bibitem[{{Plotkin} {et~al}\mbox{.}(2012){Plotkin}, {Markoff}, {Kelly},
  {K{\"o}rding}, \& {Anderson}}]{plotkin12a}
{Plotkin} R.~M., {Markoff} S., {Kelly} B.~C., {K{\"o}rding} E., {Anderson}
  S.~F., 2012, \mnras, 419, 267

\bibitem[{{Polko}, {Meier} \& {Markoff}(2013){Polko}, {Meier}, \&
  {Markoff}}]{polko13}
{Polko} P., {Meier} D.~L., {Markoff} S., 2013, \mnras, 428, 587

\bibitem[{{Polko}, {Meier} \& {Markoff}(2014){Polko}, {Meier}, \&
  {Markoff}}]{polko14}
{Polko} P., {Meier} D.~L., {Markoff} S., 2014, \mnras, 438, 959

\bibitem[{{Quataert} \& {Gruzinov}(2000)}]{quataert00}
{Quataert} E., {Gruzinov} A., 2000, \apj, 539, 809

\bibitem[{{Ratti} {et~al}\mbox{.}(2012){Ratti}, {Jonker}, {Miller-Jones},
  {Torres}, {Homan}, {Markoff}, {Tomsick}, {Kaaret}, {Wijnands}, {Gallo},
  {{\"O}zel}, {Steeghs}, \& {Fender}}]{ratti12}
{Ratti} E.~M., {Jonker} P.~G., {Miller-Jones} J.~C.~A. {et~al.}, 2012, \mnras,
  423, 2656

\bibitem[{{Rau}, {Greiner} \& {Filgas}(2011){Rau}, {Greiner}, \&
  {Filgas}}]{rau11}
{Rau} A., {Greiner} J., {Filgas} R., 2011, The Astronomer's Telegram, 3140, 1

\bibitem[{{Reis}, {Fabian} \& {Miller}(2010){Reis}, {Fabian}, \&
  {Miller}}]{reis10}
{Reis} R.~C., {Fabian} A.~C., {Miller} J.~M., 2010, \mnras, 402, 836

\bibitem[{{Remillard} \& {McClintock}(2006)}]{remillard06}
{Remillard} R.~A., {McClintock} J.~E., 2006, \araa, 44, 49

\bibitem[{{Reynolds} {et~al}\mbox{.}(2014){Reynolds}, {Reis}, {Miller},
  {Cackett}, \& {Degenaar}}]{reynolds14}
{Reynolds} M.~T., {Reis} R.~C., {Miller} J.~M., {Cackett} E.~M., {Degenaar} N.,
  2014, \mnras, 441, 3656

\bibitem[{{Rodriguez} {et~al}\mbox{.}(2007){Rodriguez}, {Cadolle Bel},
  {Tomsick}, {Corbel}, {Brocksopp}, {Paizis}, {Shaw}, \&
  {Bodaghee}}]{rodriguez07}
{Rodriguez} J., {Cadolle Bel} M., {Tomsick} J.~A., {Corbel} S., {Brocksopp} C.,
  {Paizis} A., {Shaw} S.~E., {Bodaghee} A., 2007, \apjl, 655, L97

\bibitem[{{Roming} {et~al}\mbox{.}(2005){Roming}, {Kennedy}, {Mason}, {Nousek},
  {Ahr}, {Bingham}, {Broos}, {Carter}, {Hancock}, {Huckle}, {Hunsberger},
  {Kawakami}, {Killough}, {Koch}, {McLelland}, {Smith}, {Smith}, {Soto},
  {Boyd}, {Breeveld}, {Holland}, {Ivanushkina}, {Pryzby}, {Still}, \&
  {Stock}}]{roming05}
{Roming} P.~W.~A., {Kennedy} T.~E., {Mason} K.~O. {et~al.}, 2005, \ssr, 120, 95

\bibitem[{{Russell} {et~al}\mbox{.}(2010){Russell}, {Maitra}, {Dunn}, \&
  {Markoff}}]{russell10}
{Russell} D.~M., {Maitra} D., {Dunn} R.~J.~H., {Markoff} S., 2010, \mnras, 405,
  1759

\bibitem[{{Russell} {et~al}\mbox{.}(2013){Russell}, {Markoff}, {Casella},
  {Cantrell}, {Chatterjee}, {Fender}, {Gallo}, {Gandhi}, {Homan}, {Maitra},
  {Miller-Jones}, {O'Brien}, \& {Shahbaz}}]{russell13}
{Russell} D.~M., {Markoff} S., {Casella} P. {et~al.}, 2013, \mnras, 429, 815

\bibitem[{{Rybicki} \& {Lightman}(1979)}]{rybicki79}
{Rybicki} G.~B., {Lightman} A.~P., 1979, {Radiative processes in astrophysics}

\bibitem[Serabyn et al.(1997)]{serabyn97} Serabyn, E., Carlstrom, 
J., Lay, O., et al.\ 1997, \apjl, 490, L77 

\bibitem[{{Schirmer}(2013)}]{schirmer13}
{Schirmer} M., 2013, \apjs, 209, 21

\bibitem[{{Shahbaz} {et~al}\mbox{.}(2013){Shahbaz}, {Russell}, {Zurita},
  {Casares}, {Corral-Santana}, {Dhillon}, \& {Marsh}}]{shahbaz13}
{Shahbaz} T., {Russell} D.~M., {Zurita} C., {Casares} J., {Corral-Santana}
  J.~M., {Dhillon} V.~S., {Marsh} T.~R., 2013, \mnras, 434, 2696

\bibitem[{{Sivakoff}, {Miller-Jones} \& {Krimm}(2011){Sivakoff},
  {Miller-Jones}, \& {Krimm}}]{sivakoff11}
{Sivakoff} G.~R., {Miller-Jones} J.~C.~A., {Krimm} H.~A., 2011, The
  Astronomer's Telegram, 3147, 1

\bibitem[{{Skrutskie} {et~al}\mbox{.}(2006){Skrutskie}, {Cutri}, {Stiening},
  {Weinberg}, {Schneider}, {Carpenter}, {Beichman}, {Capps}, {Chester},
  {Elias}, {Huchra}, {Liebert}, {Lonsdale}, {Monet}, {Price}, {Seitzer},
  {Jarrett}, {Kirkpatrick}, {Gizis}, {Howard}, {Evans}, {Fowler}, {Fullmer},
  {Hurt}, {Light}, {Kopan}, {Marsh}, {McCallon}, {Tam}, {Van Dyk}, \&
  {Wheelock}}]{skrutskie06}
{Skrutskie} M.~F., {Cutri} R.~M., {Stiening} R. {et~al.}, 2006, \aj, 131, 1163

\bibitem[{{Soleri} \& {Fender}(2011)}]{soleri11}
{Soleri} P., {Fender} R., 2011, \mnras, 413, 2269

\bibitem[{{Steele} {et~al}\mbox{.}(2004){Steele}, {Smith}, {Rees}, {Baker},
  {Bates}, {Bode}, {Bowman}, {Carter}, {Etherton}, {Ford}, {Fraser}, {Gomboc},
  {Lett}, {Mansfield}, {Marchant}, {Medrano-Cerda}, {Mottram}, {Raback},
  {Scott}, {Tomlinson}, \& {Zamanov}}]{steele04}
{Steele} I.~A., {Smith} R.~J., {Rees} P.~C. {et~al.}, 2004, in Society of
  Photo-Optical Instrumentation Engineers (SPIE) Conference Series, Vol. 5489,
  Ground-based Telescopes, {Oschmann} Jr. J.~M., ed., pp. 679--692

\bibitem[{{Torres} {et~al}\mbox{.}(2015){Torres}, {Jonker}, {Miller-Jones},
  {Steeghs}, {Repetto}, \& {Wu}}]{torres15}
{Torres} M.~A.~P., {Jonker} P.~G., {Miller-Jones} J.~C.~A., {Steeghs} D.,
  {Repetto} S., {Wu} J., 2015, \mnras, 450, 4292

\bibitem[{{Weng} \& {Zhang}(2015)}]{weng15}
{Weng} S.-S., {Zhang} S.-N., 2015, \mnras, 447, 486

\bibitem[{{Wright} {et~al}\mbox{.}(2010){Wright}, {Eisenhardt}, {Mainzer},
  {Ressler}, {Cutri}, {Jarrett}, {Kirkpatrick}, {Padgett}, {McMillan},
  {Skrutskie}, {Stanford}, {Cohen}, {Walker}, {Mather}, {Leisawitz}, {Gautier},
  {McLean}, {Benford}, {Lonsdale}, {Blain}, {Mendez}, {Irace}, {Duval}, {Liu},
  {Royer}, {Heinrichsen}, {Howard}, {Shannon}, {Kendall}, {Walsh}, {Larsen},
  {Cardon}, {Schick}, {Schwalm}, {Abid}, {Fabinsky}, {Naes}, \&
  {Tsai}}]{wright10}
{Wright} E.~L., {Eisenhardt} P.~R.~M., {Mainzer} A.~K. {et~al.}, 2010, \aj,
  140, 1868

\bibitem[{{Xue} \& {Cui}(2007)}]{xue07}
{Xue} Y.~Q., {Cui} W., 2007, \aap, 466, 1053

\bibitem[{{Yuan} \& {Cui}(2005)}]{yuan05a}
{Yuan} F., {Cui} W., 2005, \apj, 629, 408

\bibitem[{{Yuan}, {Cui} \& {Narayan}(2005){Yuan}, {Cui}, \& {Narayan}}]{yuan05}
{Yuan} F., {Cui} W., {Narayan} R., 2005, \apj, 620, 905

\end{thebibliography}

\end{document}